\documentclass[
    ,final            
  ]
  {aipproc}

\layoutstyle{6x9}

\usepackage{epsf,graphicx}
\newcommand{\Msun}{\mbox{$M_{\odot}$}}

\begin{document}

\title{Linear Spectropolarimetry and the Circumstellar Media of Young and Massive Stars}


\author{Jorick S. Vink}{
  address={Armagh Observatory, College Hill, BT61 9DG, Armagh, Northern Ireland, UK}
}

\begin{abstract}
Linear spectropolarimetry is a powerful tool to probe circumstellar structures on spatial scales that cannot yet be achieved
through direct imaging. In this review I discuss the role that emission-line polarimetry can play in constraining geometrical and 
physical properties of a wide range of 
circumstellar environments, varying from the accretion disks around pre-main sequence T Tauri and Herbig Ae/Be stars, to the issue 
of stellar wind clumping, and the aspherical outflows from the massive star progenitors of supernovae and long gamma-ray bursts 
at low metallicity. 
\end{abstract}

\maketitle


\section{Introduction}

Massive stars and pre-main sequence (PMS) T Tauri and Herbig Ae/Be stars 
have comparable astrophysical ages, of a few million years. 
They also share the ubiquitous presence of circumstellar media (CSM), 
involving disks and winds responsible for their emission lines, and 
which can be studied using similar tools.
Key issues in both topics of star formation and massive-star evolution involve 
angular momentum physics, which remains poorly understood.

Future instruments including the JWST and the E-ELT focus on questions relating to 
the early Cosmos, involving the epoch of reionization, as well as 
local star formation. However, pivotal properties 
of both PMS and low-metallicity massive stars remain largely unknown. 
In the following I argue that linear polarimetry can play a relevant role in removing these 
uncertainties.

With the detection of long-duration gamma-ray bursts (GRBs) at redshifts up to $\sim$9-10, 
just a few hundred million years after the Big Bang, and the potential detection of super-luminous
supernovae (SNe) out to redshifts $>$4 \cite{quimby11} 
with instruments such as PTF, Pan-STARRS, and LSST, we urgently need to understand 
the progenitors of these cosmic explosions. Currently we do not. 

We know that stars up to $\sim$15\Msun\ produce hydrogen (H) rich type II SNe \cite{smartt09}. 
However, the massive progenitors of all other core-collapse 
SNe types, whether involving H-poor Ibc SNe, interacting IIn SNe \cite{hoffman08}, or 
even pair-instability SNe 
(PISNs), where the entire star is disrupted; all metals are released; and no remnant is left, remain as yet 
elusive. What is clear is that their evolution towards collapse is driven by mass loss and rotation, which are highly 
intertwined. 

There have been several suggestions that rotation affects both the strength and  
the latitudinal dependence of their outflows, but in turn stellar winds are thought to 
remove significant amounts of angular momentum, possibly down to masses 
as low as $\sim$10\Msun\ \cite{vink10}. 
Whether mass loss is latitude dependent, occurring from the 
pole \citep{owocki96}, maintaining rapid rotation \citep{mm07}, or 
from the equator \citep{bc93,pel00}, subject to loss of spin, remains 
an open question. 

Given the requirements on spatial resolution in low $Z$ extragalactic environments, where 
both GRBs \cite{woos06} and PISNs \cite{langer07} have been favoured, because of 
lower mass-loss rates \cite{vdk05}, linear polarimetry may be the only way to resolve 
these issues.

Questions such as whether certain massive stars in particular redshift bins may make 
GRBs will ultimately rely on the initial conditions, 
determined by the process of the star's formation. 
Whilst many aspects of low-mass star formation seem well established in that there is
at least a paradigm describing the physical properties of the optically visible 
T Tauri stars: ``magnetospheric accretion'', the situation for the 
intermediate mass (2-15\Msun) Herbig Ae/Be stars becomes more patchy, whilst for 
the most massive stars it is not even clear as to whether they might form via disk accretion 
at all. 

\section{The tool of spectropolarimetry}

The basic idea of linear spectropolarimetry is very simple. It 
is largely based on the premise that free electrons in an extended 
ionized CSM scatter the continuum radiation from the central star,  
revealing a certain amount of linear polarization. 
If the projected electron distribution is perfectly 
circular, e.g. when the CSM is spherically symmetric 
or when a CS disk is observed face-on, the linear 
Stokes vectors $Q$ and $U$ cancel, and no polarization 
is observed (as long as the object is spatially unresolved). 
If the geometry is not circular but involves  
an inclined CS disk this is expected to result in some net 
continuum polarization.  

\begin{figure}
\resizebox{0.5\columnwidth}{!}{\includegraphics{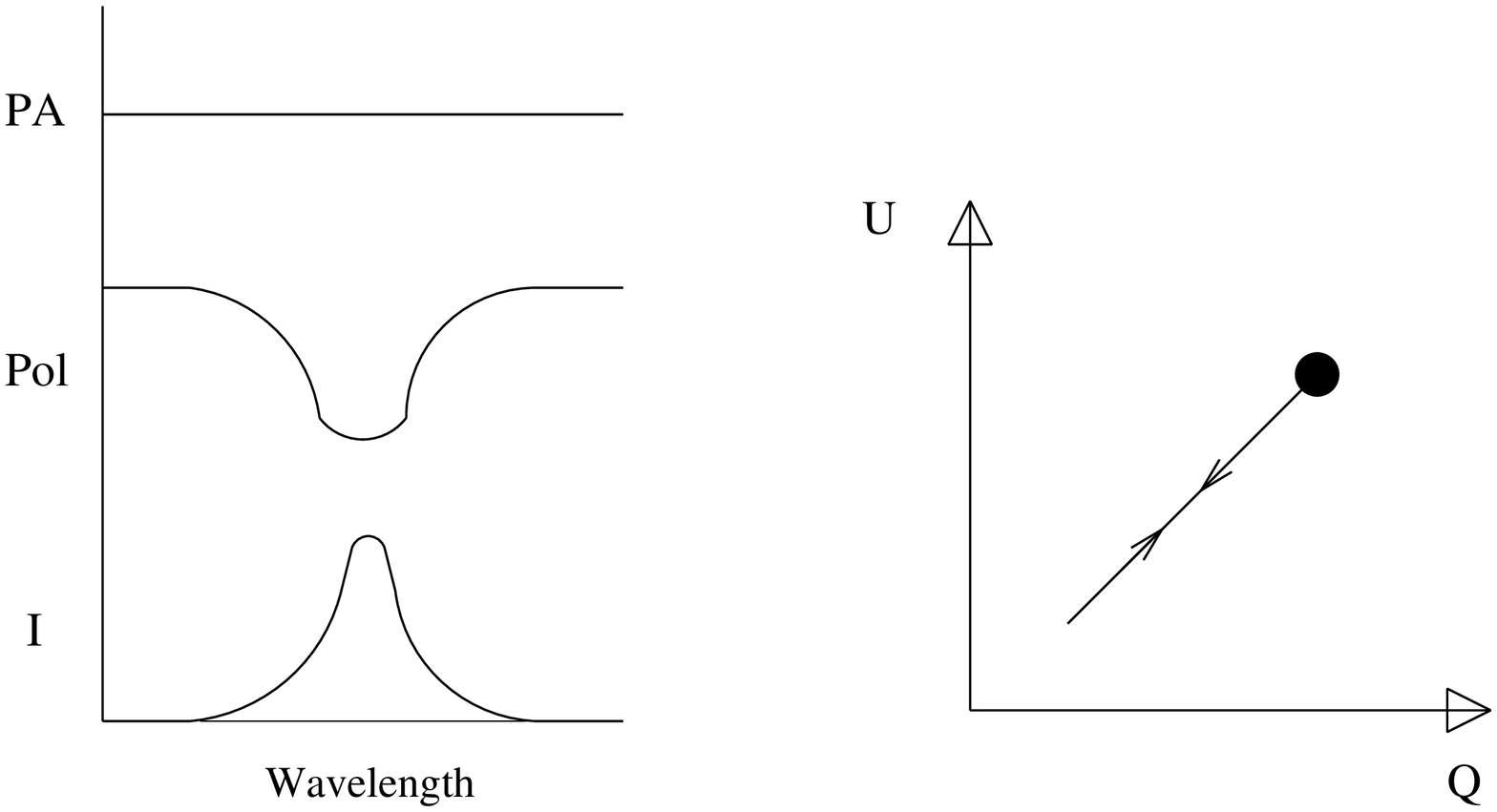}}
\resizebox{0.5\columnwidth}{!}{\includegraphics{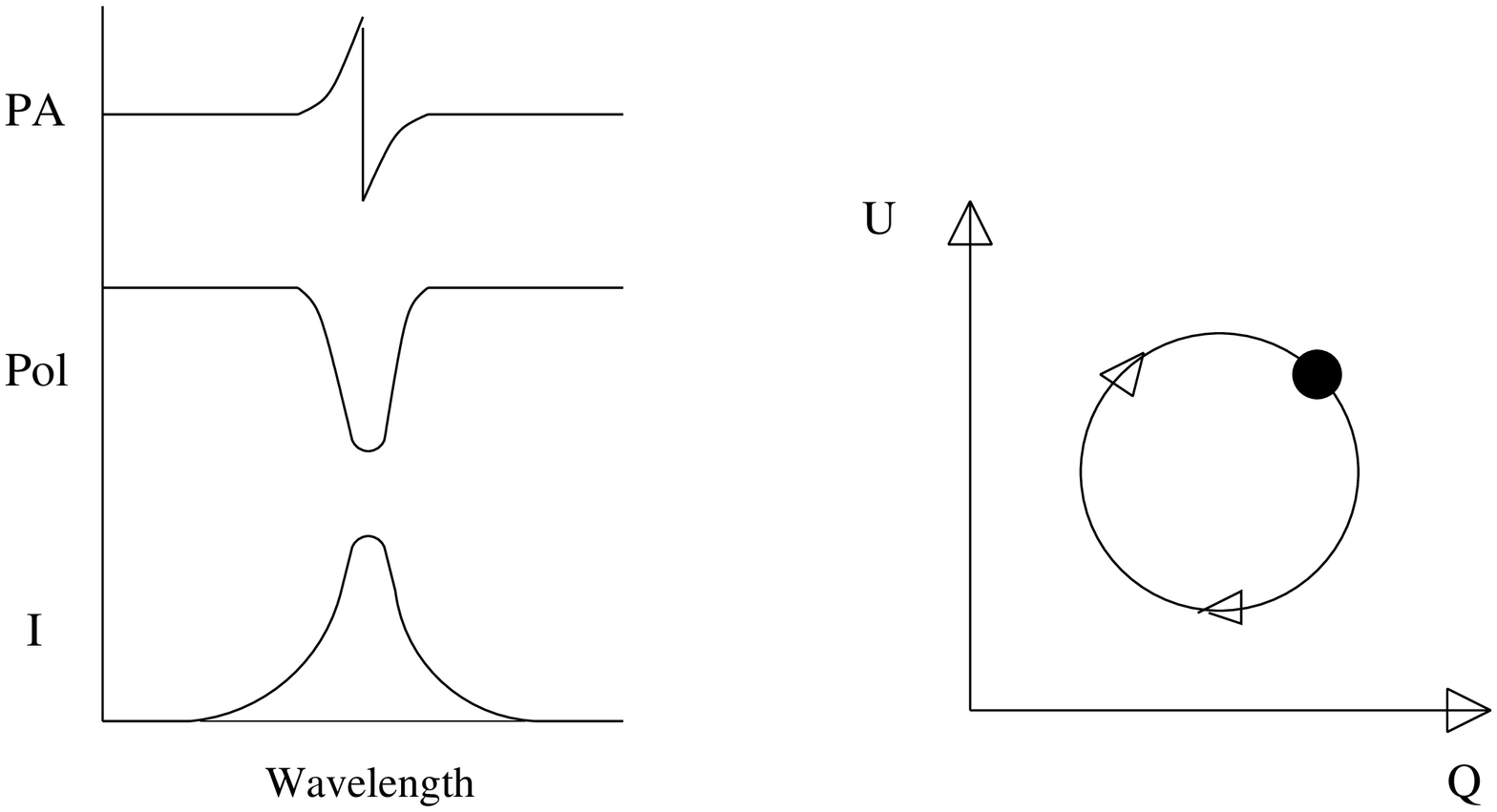}}
\caption{Cartoons representing line {\it depolarization} (left hand side) and compact line emission
scattered off a rotating disk (right hand side) as triplots and $QU$ diagrams. Stokes $I$
profiles are shown in the lower triplot panels, \% Pol in the middle panels, and 
the position angles (PAs) are given in the upper triplot panels. 
Line depolarization is as broad as the Stokes $I$ emission, while the {\it line} 
polarization is narrow by comparison. Depolarization translates 
into $QU$ space as a linear excursion (left hand side), 
whilst a {\it line} polarization PA flip is associated with a $QU$ loop (right hand side).}
\label{cartoons}     
\end{figure}

One of the advantages of spectropolarimetry over continuum polarimetry is 
that one can perform differential polarimetry between a spectral line and the 
continuum {\it independent} of any interstellar or instrumental 
polarization. The H$\alpha$ depolarization ``line effect'' utilizes the expectation 
that hydrogen recombination lines arise over a much larger volume than the 
continuum and becomes {\it de}polarized (see the left hand side of 
Fig.\,\ref{cartoons}). 
Depolarization immediately indicates the presence (or absence) of 
aspherical geometries, such as disks, on spatial scales that cannot be 
imaged with the world's largest telescopes. 

The basic idea of the technique was explored in the 1970s by e.g. Poeckert
\& Marlborough \cite{pm76} who employed narrow-band filters to show that
Be stars have CS disks as around 55\% of their objects showed the 
depolarization line effect. 
It took another couple of decades before interferometry 
\citep{doug} could confirm
these early findings. Interestingly, in a recent 
study of peculiar O stars, \citep{vink09} 
did not find evidence for the presence of disks in Oe stars -- the alleged 
counterparts of classical Be stars -- although the first detection of a line effect 
in an Oe star (HD\,45314) was reported. 

In general we divide the polarimetric data into bins 
corresponding to 0.1\% polarization, the typical error bar (although 
the numbers from photon statistics are at least a factor 10 better). 
We also present the data in $QU$ diagrams. 
For the case of line depolarization this translates into 
a linear excursion from the cluster of points that represents the continuum 
($P^2 = Q^2 + U^2$) with the excursion showing the trend 
when the polarization moves in and out of line center.  

A most relevant quantity involves the detection limit, which is inversely 
dependent on the signal-to-noise ratio (SNR) and the contrast of the emission line
to the continuum. 
The detection limit $\Delta P_{\rm limit}$ can be represented by:

\begin{equation}
\Delta P_{\rm limit} (\%) = \frac{100}{SNR} \times \frac{l/c}{l/c-1}
\label{eq:pint}
\end{equation} 

\noindent where $l/c$ refers to the line-to-continuum contrast. 
This detection limit is most useful for objects with strong emission lines, such 
as H$\alpha$ emission in Luminous Blue Variables (LBVs; \cite{davies05}), where the emission
completely overwhelms underlying photospheric absorption. 
In general, we aim for an SNR in the continuum of 1000, corresponding
to changes in the amount of linear polarization of 0.1\%. 
We should be able to infer asymmetry degrees in the form of 
equator/pole density ratios, $\rho_{\rm eq}/\rho_{\rm pole}$ of $\sim$1.25, or larger 
\citep{harries00}, with some small additional dependence on the shape and 
inclination of the disk \citep{brown}.

Most of the linear line polarimetry work of the last two decades has indeed concerned
line depolarization, but in the following we will see that in some cases there is 
evidence for intrinsic {\it line} polarization, predicted by \cite{wood93} and 
found observationally in PMS T Tauri and Herbig Ae/Be stars by our group 
\citep{vink02,vink03,vink05a,vink05b,mott}.
In such cases line photons are thought to originate from a {\it compact} source, e.g.
as a result of (magnetospheric) accretion. These compact photons are scattered off a rotating 
disk, leading to a flip in the position angle (PA), and resulting in a rounded loop (rather than a linear excursion) 
in the $QU$ diagram (sketched on the right hand side of Fig.\,\ref{cartoons}).

\section{Pre-main sequence stars and their accretion disks}

\begin{figure}
\centering
\includegraphics[height=8cm]{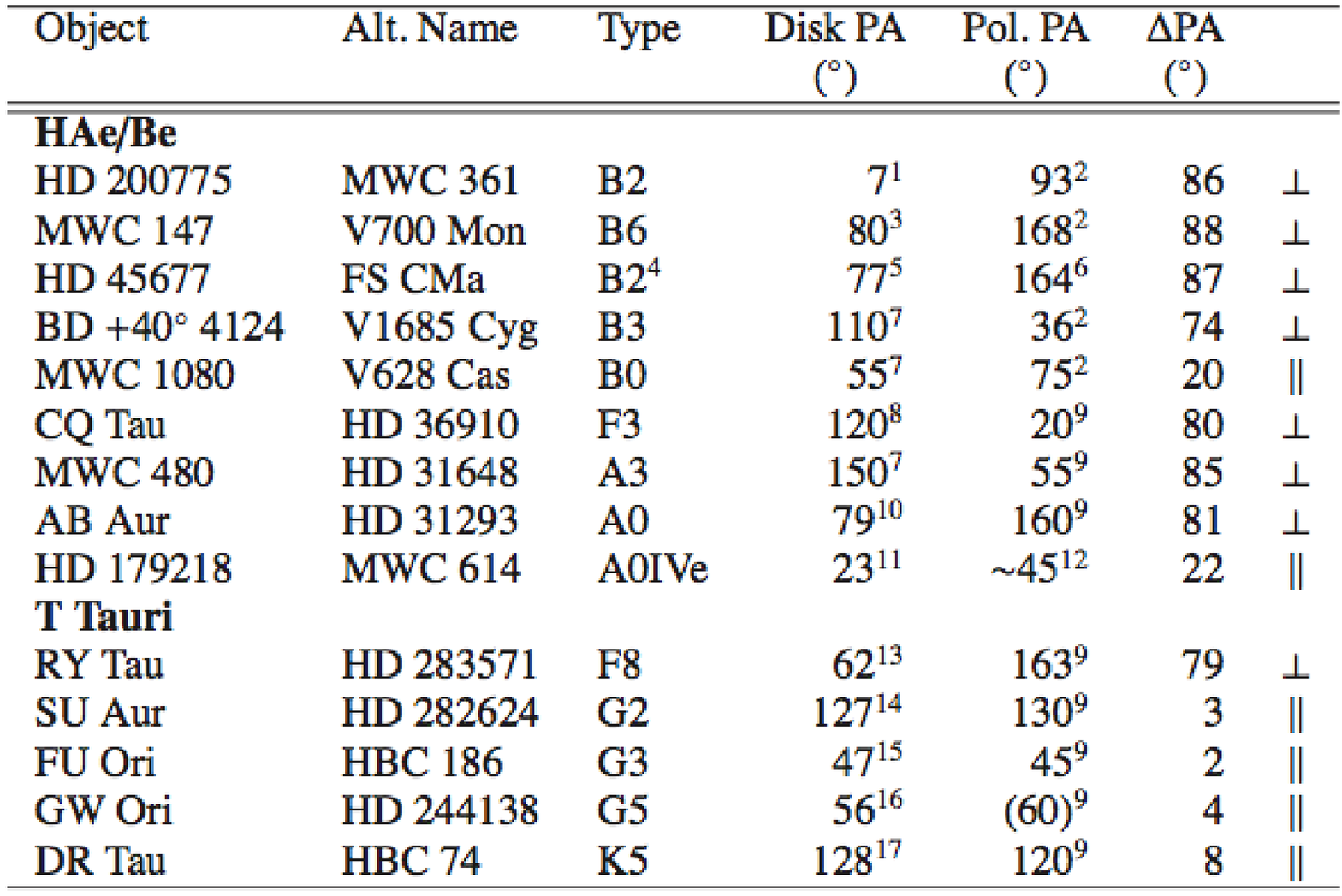}
\caption{Comparison of polarization PAs versus imaged disk PAs for a selection 
of binary PMS from \cite{wheel11}. See \cite{vink05b} for a larger sample of polarization PAs.}
\label{tab}     
\end{figure}

\begin{figure}
\centering
\includegraphics[height=7.15cm]{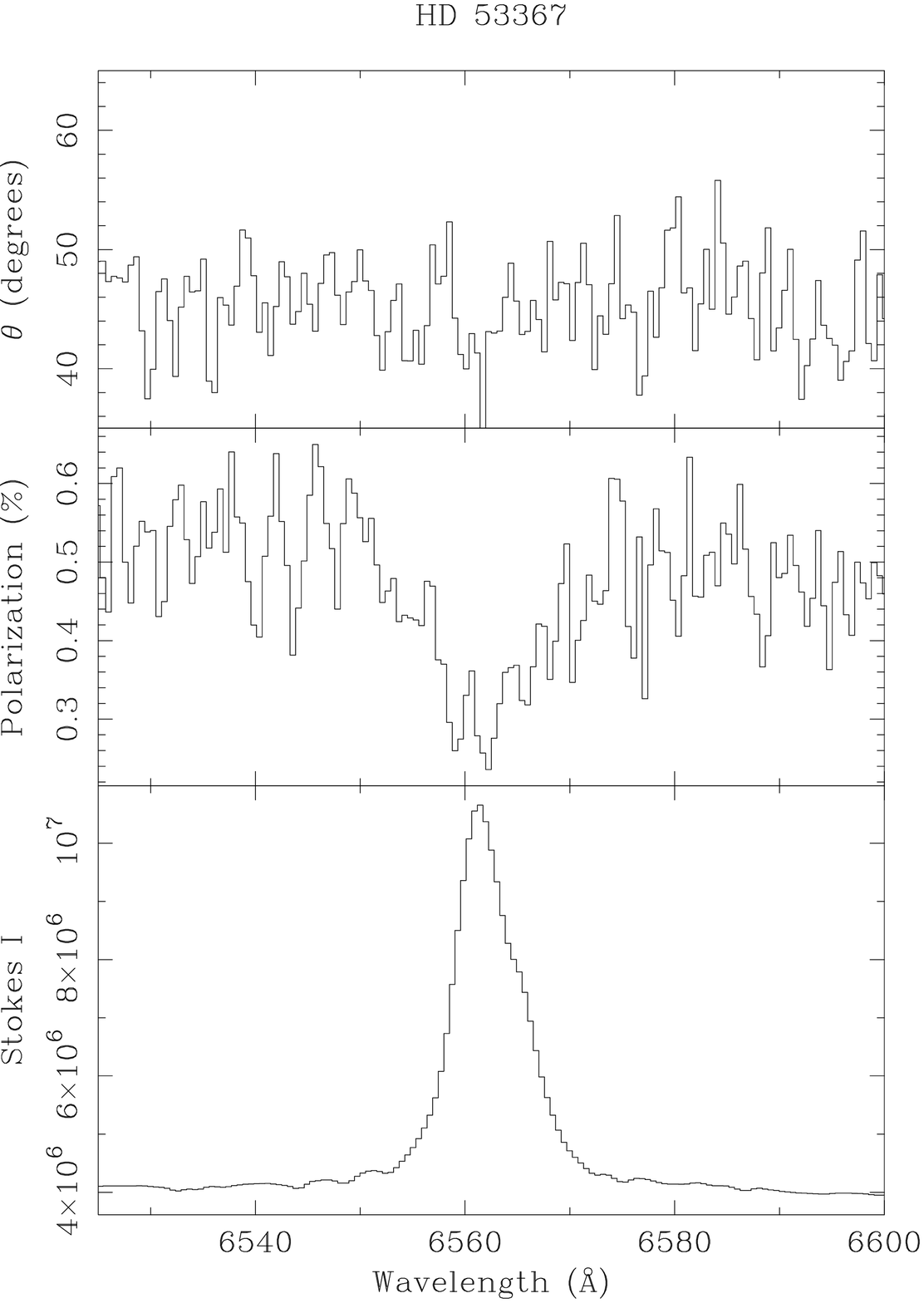}
\includegraphics[height=7.15cm]{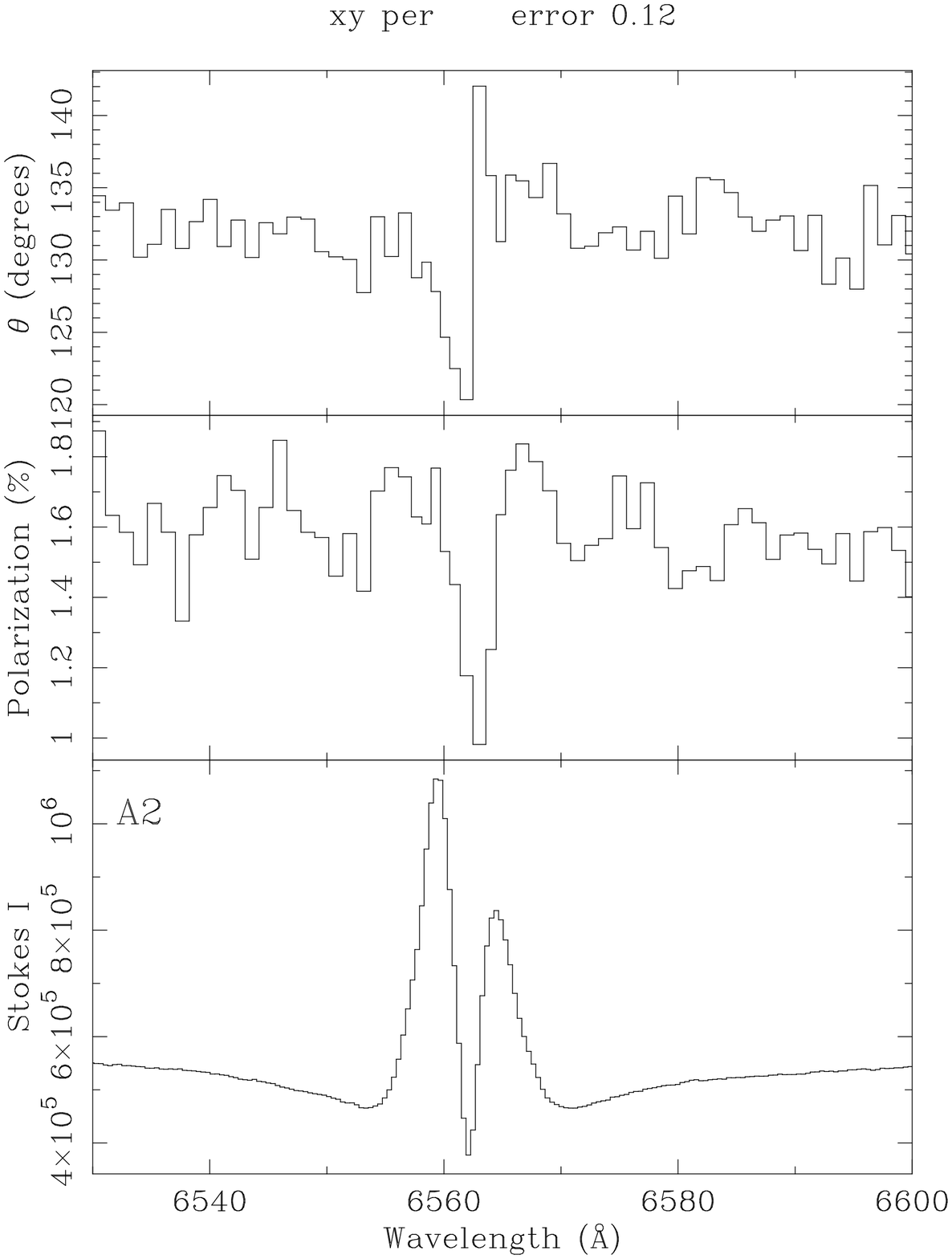}
\includegraphics[height=7.15cm]{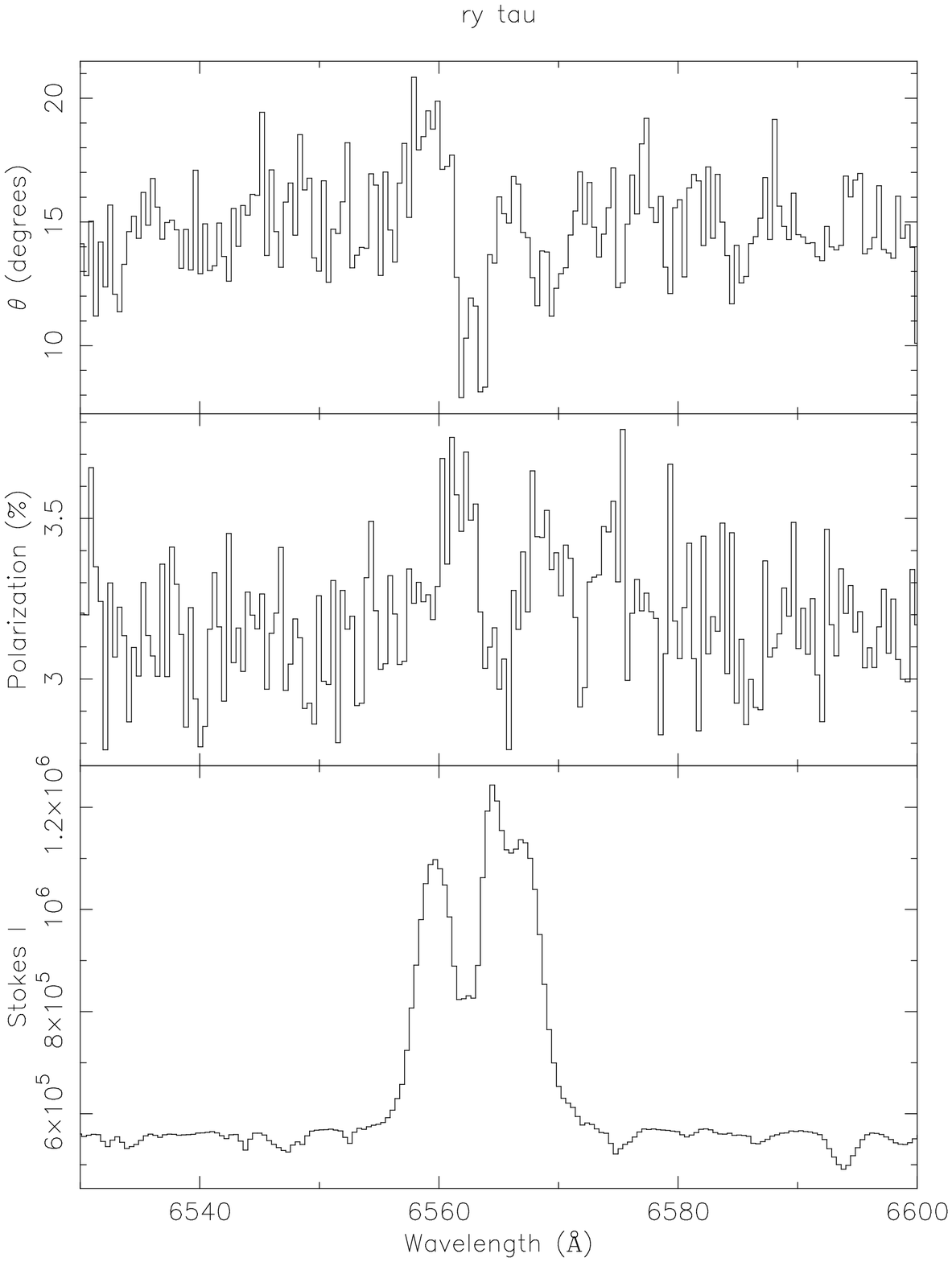}
\caption{H$\alpha$ spectropolarimetry of a Herbig Be object (left), a Herbig Ae
star (middle) and a T Tauri star (right). As discussed in the text,
the Herbig Ae stars show larger spectropolarimetric resemblance
to T Tauri stars than to Herbig Be stars.}
\label{three}     
\end{figure}

Over the last decade we have surveyed PMS Herbig Ae/Be stars and T Tauri stars
mostly using the 4m William Herschel Telescope (WHT). These 
studies involved roughly equal numbers (of over 10) for each subgroup consisting of  
$\sim$1\Msun\ T Tauri stars, $\sim$2-3\Msun\ Herbig Ae stars and 10-15\Msun\ early 
Herbig Be stars. Typical examples can be found in Fig.\,\ref{three}. 
The PA angles derived have been compared to those obtained from other techniques and 
generally show great consistency (see e.g. Fig.\,\ref{tab} taken from \cite{wheel11,vink05b}. 

\subsection{Herbig Ae/Be and T Tauri PMS data}

\begin{figure}
\centering
\includegraphics[height=8cm]{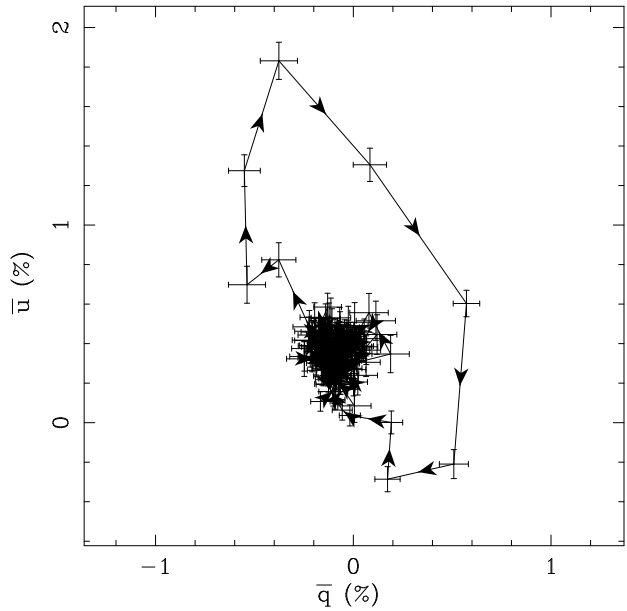}
\caption{A loop in the $QU$ diagram of the Herbig Ae star MWC\,480.}
\label{qu}     
\end{figure}

These data may provide insight into the accretion geometry 
as a function of stellar mass. One might perhaps expect the physics 
of radiation pressure to dominate the high-mass range, whilst the effects of 
stellar magnetism could dominate the later spectral types. Furthermore, there might be
a {\it transition} in accretion physics 
between the T Tauri and the Herbig Ae/Be stars, as this is the location in the stellar
Hertzsprung-Russell diagram where there is a boundary 
between the early-type OBA stars with their radiative outer layers, and the 
convective mantles of the later spectral types. 

Intriguingly, the linear spectropolarimetry data reveal a story that is very different: 
the Herbig Ae stars resemble the T Tauri data and have little in common with the early 
Herbig Be stars. The Herbig Be stars show line 
depolarization in 7/12 cases, which is consistent with the presence of 
small-scale electron scattering disks around {\it all} of them. This is 
a very similar incidence rate as was found for the classical Be stars, and these data are 
consistent with direct disk accretion for the more massive stars \citep{od99,vink02}.

The question is how to interpret the PA flips seen in the Herbig Ae and T Tauri stars. 
When we plot the data in $QU$ space we find $QU$ loops (e.g. Fig.\,\ref{qu}).

\subsection{Monte Carlo modelling}

\begin{figure}
\resizebox{0.5\columnwidth}{!}{\includegraphics{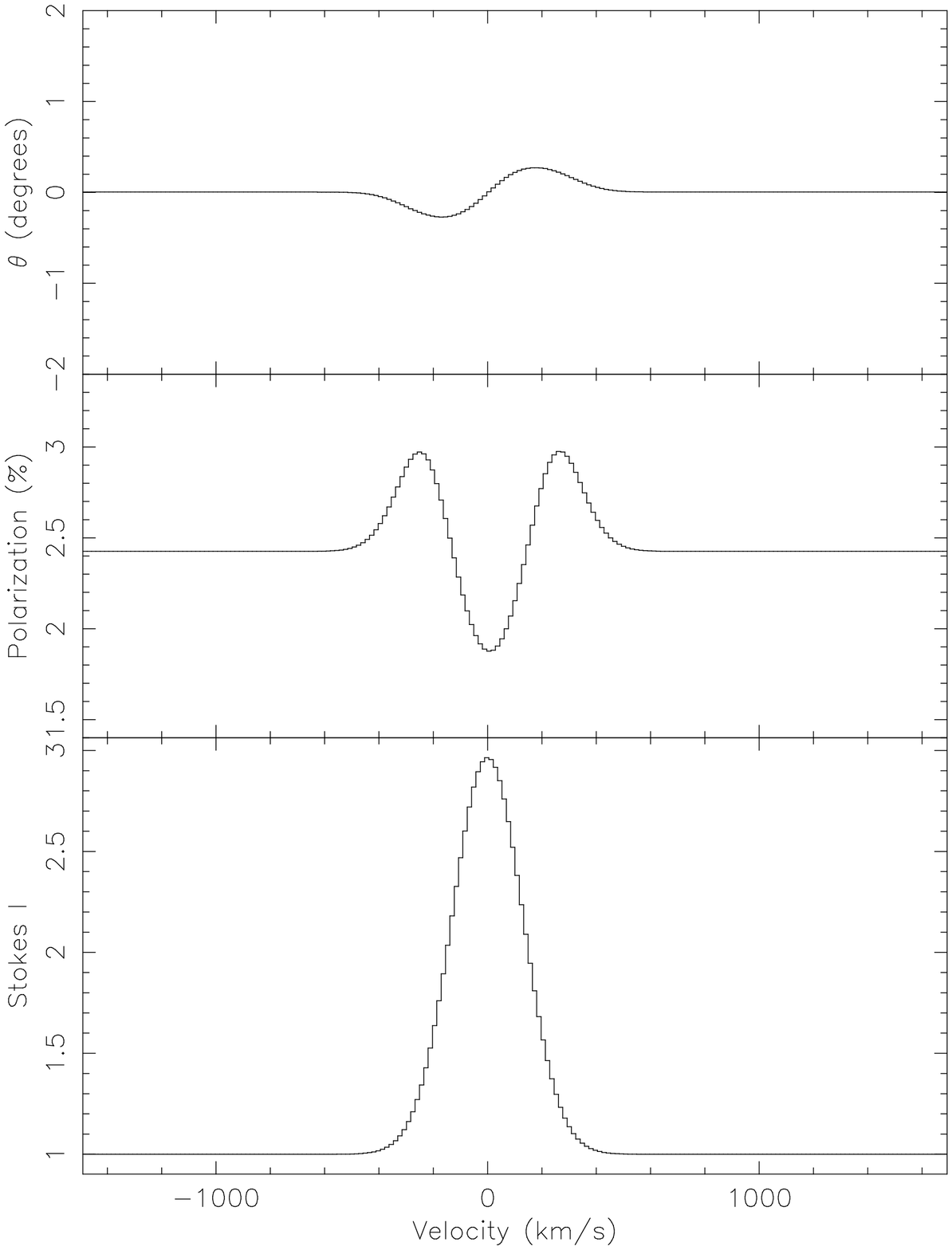}}
\resizebox{0.5\columnwidth}{!}{\includegraphics{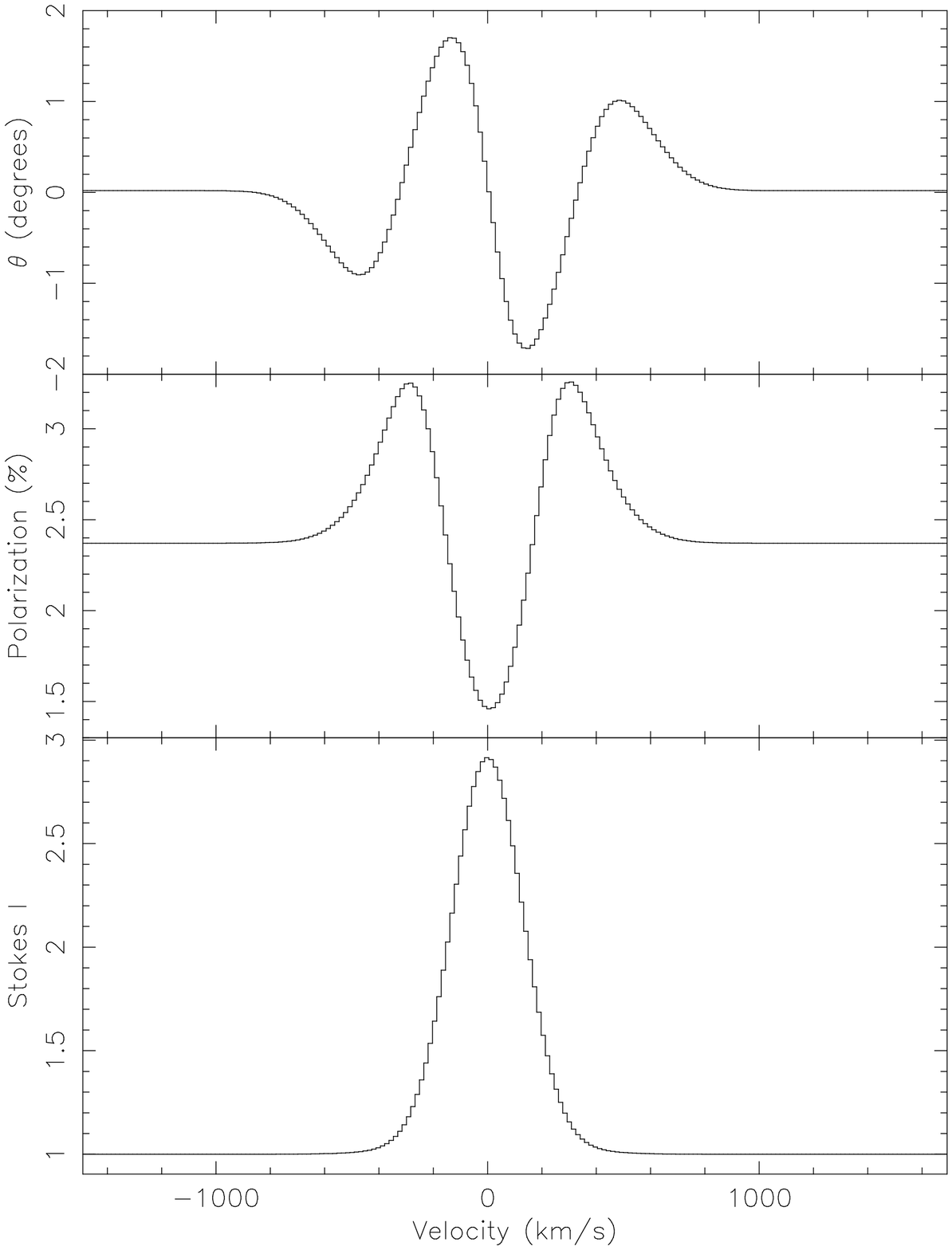}}
\caption{Monte Carlo line polarimetry predictions for the case 
of a disk with an inner hole (left hand side) 
and without an inner hole -- a result of the finite size of the 
star (right hand side). From \cite{vink05a}.}
\label{mc}     
\end{figure}

Motivated by the almost {\it ubiquitous} incidence of $QU$ loops in T Tauri and Herbig Ae stars 
(\cite{vink02,vink03,vink05b} but see also \cite{harrington09} for alternative findings/interpretations), we 
decided to develop numerical polarization models of line emission scattered off 
Keplerian rotating disks \citep{vink05a} using the 3D Monte Carlo code {\sc torus} \citep{harries00}, both 
with and without a disk inner hole. 
Figure\,\ref{mc} shows a marked difference between scattering off a disk that reaches the stellar photosphere 
(right hand side), and a disk with a significant inner hole (left hand side). 
The single PA flip on the left-hand side is similar to that predicted analytically \citep{wood93}, but 
the double PA flip on the right-hand side -- associated with the undisrupted disk -- came as a surprise at the time. 
The effect is the result of the geometrically correct treatment of the finite-sized stars that interacts with the disk's rotational 
velocity field. 
Our numerical models demonstrate the diagnostic potential of {\it line} polarimetry (as opposed to simple depolarization, where 
no velocity information can be obtained) in determining not only the disk inclination, but also the size of 
disk inner holes. 
As far as we are aware linear {\it line} polarimetry is as yet the {\it only} method 
capable of determining disk hole sizes on the required spatial scales. 

\section{Disks and clumps of massive stars}

\begin{figure}
\includegraphics[height=8cm]{agcar_tri2}
\includegraphics[height=8cm]{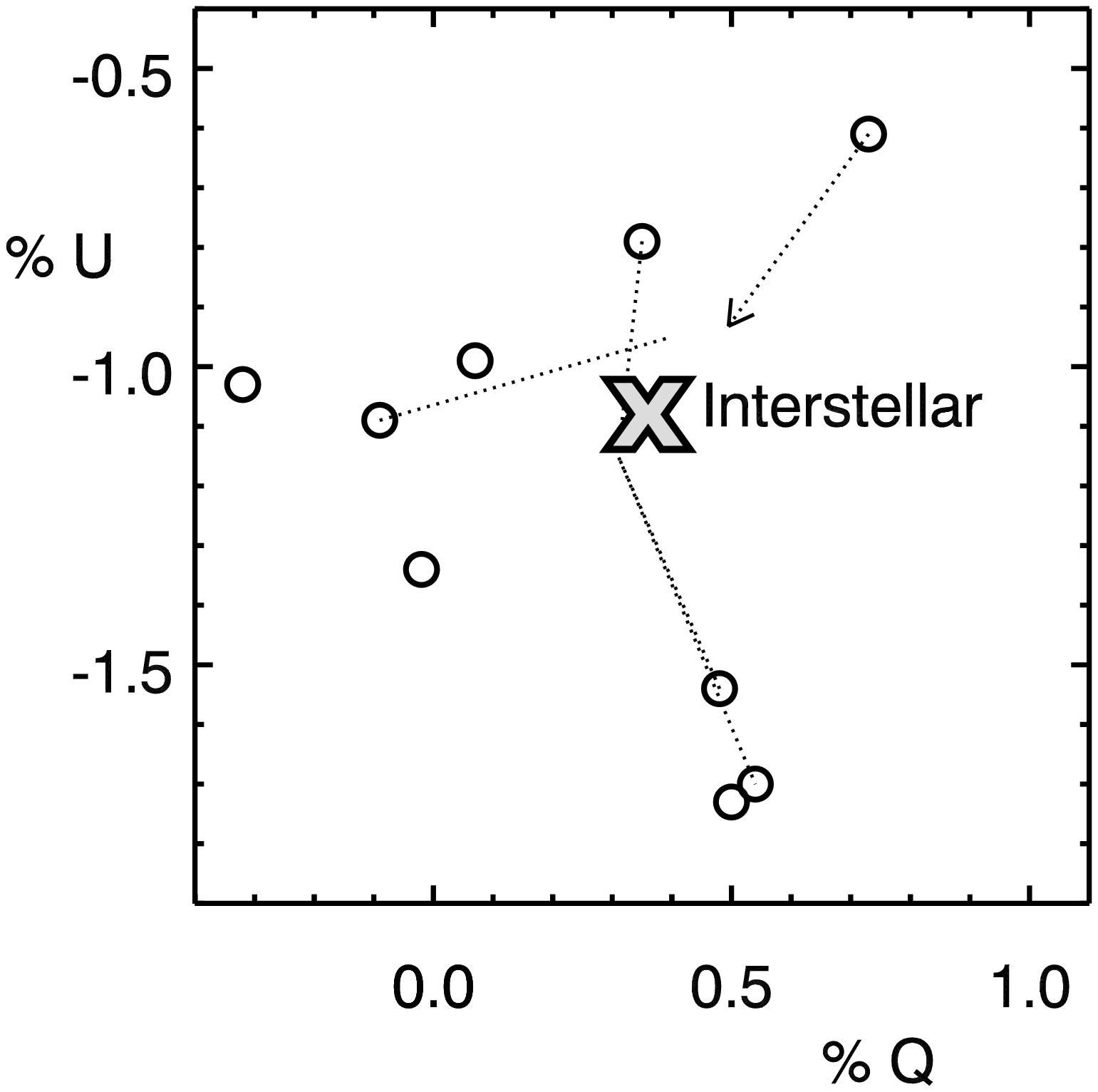}
\caption{H$\alpha$ polarimetry of the Luminous Blue Variable AG\,Car. 
The triplot on the left hand side reveals line depolarization. 
The large cross on the right hand side denotes the measured polarization at line center 
(constant with time), whilst the open circles represent the continuum measurements 
that vary with time, indicating wind clumping. See \cite{davies05,davies07} for details.}
\label{agcar}     
\end{figure}

Line depolarization has been observed in a plethora of massive stars, including 
B[e] supergiants (e.g. \cite{od99,per09}), post Red 
Supergiants \citep{patel08}, and LBVs (e.g. \cite{schulte94}). In all these 
cases the incidence rate of ``line effects'' appears to be consistent with the Be star 
results, i.e. 50-60\%. Furthermore, the measured PA of the line-effect stars shows great 
consistency with PA constraints from other techniques, providing further proof  
that the tool is capable of discovering and constraining CS disks. 

\cite{davies05} performed a spectropolarimetry survey of LBVs in the Galaxy and Magellanic Clouds and 
found some surprising results. At first sight the results suggested the presence of disks (or equatorial outflows), 
as the incidence rate of line effects was inferred to be $>$50\%. 
This is notably higher than that of their evolutionary 
neighbour O and Wolf-Rayet (WR) stars, with incidence rates of $<$25\% 
\citep{vink09} and $\sim$15\% \citep{harries98} respectively. 
However, when Davies et al. \citep{davies05} plotted the results of AG\,Car in a $QU$ diagram (see Fig.\,\ref{agcar}) they
noticed that the level of polarization varied with time, which they interpreted as the manifestation
of wind clumping. Subsequent modelling by \cite{davies07}, \cite{li09}, and Townsend (Poster this meeting) 
shows how time-variable linear polarization might become a powerful tool to constrain clump sizes and numbers. 
These constraints have already been employed in theoretical studies regarding the origin of wind 
clumping \citep{cantiello09} and the effects of wind clumping on predicted mass-loss 
rates \citep{muijres11}.

\section{The geometry of Wolf-Rayet stars at low metallicity 
and the long GRB connection}

Evidence has been accumulating that long GRBs are associated with the deaths 
of massive stars at low metallicity $Z$ \citep{woos06}.
The next piece of the puzzle is to constrain the progenitors of 
these explosive events. 
The currently most popular model is the so-called ``collapsar model'', where 
a rapidly rotating compact star collapses to a black hole. This progenitor star is likely a 
hydrogen-free WR star, but it is currently
unclear whether such an object is the result of single star or binary evolution.
In both scenarios the crucial aspect of low $Z$ is the reduced amount of angular 
momentum loss due to weaker stellar winds at low metal content. 
A recent breakthrough in the metallicity dependence of stellar 
winds from massive stars has been the finding that WR winds 
are expected to scale with the {\it iron} content (Fe) of the host 
galaxy \citep{vdk05}, and {\it not} on self-enrichment of metals such as carbon, as was 
generally assumed previously

Although it has become more firmly established that there is a low $Z$ bias in the occurrence 
of long GRBs, what is less clear is whether low $Z$ is an {\it absolute} requirement. 
With a bias in $Z$ that is low enough, one might envision a situation in which {\it all} massive single  
stars below a certain threshold $Z$ could make GRBs. 
Alternatively, the small fraction of GRB supernovae might require special circumstances 
to ensure that only a small
fraction of core-collapse supernovae occur in conjunction with a long GRB.

\begin{figure}
\centering
\includegraphics[height=8cm]{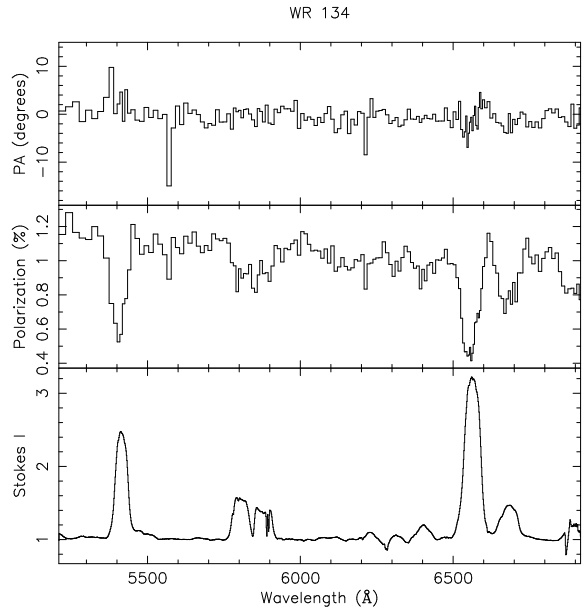}
\caption{WHT data of the Galactic Wolf-Rayet star WR 134. 
The data are taken from \cite{harries98}.}
\label{wr}     
\end{figure}

We decided to test the physical criteria of GRB progenitor models empirically using ESO's 
FORS spectropolarimeter on the Very Large Telescope (VLT). 
If the WR mass-loss metallicity dependence and the subsequent inhibition
of angular momentum removal are the {\it key} to explain the high occurrence
of GRBs at low $Z$, WR stars in the Magellanic Clouds (MCs) should on average be spinning faster than those
in the Galaxy. Vink \cite{vink07} therefore performed a linear spectropolarimetry survey of WR stars
in the low $Z$ environment of the LMC and found that $\sim$ 15\% of LMC WR stars
might show the sign of rapid rotation, as only 2 out of 13 of them show a significant amount of linear
polarization. This incidence rate equals that of the Galactic WR survey by Harries et al. (1998).
The LMC data presented in \cite{vink07} may either suggest that the metal content of the LMC is 
high enough for the WR winds to remove the angular momentum, and single star progenitors may be 
constrained to an upper metallicity of that of the LMC at $\sim$50\% solar, or alternatively, rapid 
rotation of WR stars may only be achieved for objects that are the products of a special kind of (e.g. binary) evolution. 

We emphasize that thus far, the best-fitting results (using both Kolmogorov-Smirnov tests and Monte Carlo simulations) of the Harries
et al. (see Fig.\,\ref{wr} for an example) sample were obtained if the 
majority of Galactic objects were spherically symmetric slow rotators, whilst 
the 15-20\% minority represent the more rapid rotators with large intrinsic polarizations (with 
values exceeding 0.3\%). Harries et al. therefore concluded that the inferred axi-symmetries are only present  
for the most rapidly rotating WR stars. 

\section{Summary}

In this review I have attempted to show that 
linear spectropolarimetry can become a powerful tool to study the geometries 
and sizes of circumstellar disks and clumps. Moreover:

\begin{itemize}
\item
Herbig Ae/Be stars were found to possess accretion disks on the smallest spatial scales. 
\item
Contrary to common wisdom, a transition in the HR diagram was found between the Herbig Ae and the Herbig Be stars. 
The Herbig Be data are consistent with disk accretion at high stellar masses, whilst the Ae data are more 
reminiscent of T Tauri physics.
\item
Time-variable polarimetry data suggests that wind clumping rather than the presence of 
disks is responsible for the high levels of polarization in LBVs.   
\item
Wolf-Rayet stars are found to be equally spherically symmetric in the LMC at half solar metallicity 
than in the Galaxy at solar metallicity. 
\end{itemize}

\begin{theacknowledgments}
I am extremely grateful to Rico Ignace and the other organizers fur putting together such a 
great conference in a perfect location, as well as for their generous support. 
I am also indebted to my senior and junior specpol collaborators: 
Janet Drew, Tim Harries, Rene Oudmaijer, Ben Davies, Joe Mottram, Mitesh Patel, and Hugh Wheelwright.
\end{theacknowledgments}

\bibliographystyle{aipprocl} 

\begin{thebibliography}{9}

\bibitem{quimby11}
R.~M.~Quimby, et al., \emph{Nature} \textbf{474}, 487--489 (2011).

\bibitem{smartt09}
S.~J.~Smartt, et al., \emph{MNRAS} \textbf{395}, 1409--1437 (2009).

\bibitem{hoffman08}
J.~L.~Hoffman, et al.,  \emph{ApJ} \textbf{688}, 1186--1209 (2008).

\bibitem{vink10}
J.~S.~Vink, et al., \emph{A\&A} \textbf{512}, L7$+$ (2010).

\bibitem{owocki96}
S.~P.~Owocki, et al., \emph{ApJ} \textbf{472}, L115$+$ (1996).

\bibitem{mm07}
G.~Meynet, and A.~Maeder, \emph{A\&A} \textbf{464} L11--L15 (2007).

\bibitem{bc93}
J.~E.~Bjorkman, and J.~P.~Cassinelli,  \emph{ApJ} \textbf{409} 429-449 (1993).

\bibitem{pel00}
F.~I.~Pelupessy, et al., \emph{A\&A} \textbf{359} 695--706 (2000).

\bibitem{woos06}
S.~E.~Woosley, and A.~Heger, \emph{ApJ} \textbf{637} 914--921 (2006).

\bibitem{langer07}
N.~Langer, et al., \emph{A\&A} \textbf{475} L19--L23 (2007).

\bibitem{vdk05}
J.~S.~Vink, and A.~de Koter, \emph{A\&A} \textbf{442} 587--596 (2005).

\bibitem{pm76}
R.~Poeckert, and J.~M.~Marlborough, \emph{ApJ} \textbf{206} 182--195 (1976).

\bibitem{doug}
S.~M.~Dougherty, and A.~R.~Taylor, \emph{Nature} \textbf{359} 808--810 (1992).

\bibitem{vink09}
J.~S.~Vink, et al., \emph{A\&A} \textbf{505} 743--753 (2009).

\bibitem{davies05}
B.~Davies, et al., \emph{A\&A} \textbf{439} 1107--1125 (2005).

\bibitem{harries00}
T.~J.~Harries, \emph{MNRAS} \textbf{315} 722--734 (2000).

\bibitem{brown}
J.~C.~Brown, and I.~S.~McLean, \emph{A\&A} \textbf{57} 141--$+$ (1977).

\bibitem{wood93}
K.~Wood, et al., \emph{A\&A} \textbf{271} 492--$+$ (1993).

\bibitem{vink02}
J.~S.~Vink, et al., \emph{MNRAS} \textbf{337} 356--368 (2002).

\bibitem{vink03}
J.~S.~Vink, et al., \emph{A\&A} \textbf{406} 703--707 (2003).

\bibitem{vink05a}
J.~S.~Vink, et al., \emph{A\&A} \textbf{430} 213--222 (2005a).

\bibitem{vink05b}
J.~S.~Vink, et al., \emph{MNRAS} \textbf{359} 1049--1064 (2005b).

\bibitem{mott}
J.~C.~Mottram, \emph{MNRAS} \textbf{377} 1363--1374 (2007).

\bibitem{wheel11}
H.~E.~Wheelwright, et al., \emph{A\&A} \textbf{532} A28$+$ (2011).

\bibitem{od99}
R.~D.~Oudmaijer, and J.~E.~Drew, \emph{MNRAS} \textbf{305} 166--180 (1999).

\bibitem{harrington09}
D.~M.~Harrington, and J.~R.~Kuhn, \emph{ApJ} \textbf{667} L89--L92 (2007).

\bibitem{per09}
A.~Pereyra, et al., \emph{A\&A} \textbf{508} 1337--1341 (2009).

\bibitem{patel08}
M.~Patel, et al., \emph{MNRAS} \textbf{385} 967--978 (2008).

\bibitem{schulte94}
R.~E.~Schulte-Ladbeck, et al., \emph{ApJ} \textbf{429} 846--856 (1994).

\bibitem{harries98}
T.~J.~Harries, et al., \emph{MNRAS} \textbf{296} 1072--1088 (1998).

\bibitem{davies07}
B.~Davies, et al., \emph{A\&A} \textbf{469} 1045--1056 (2007).

\bibitem{li09}
Q.~-K.~Li, et al., \emph{RAA} \textbf{9} 558--576 (2009)

\bibitem{cantiello09}
M.~Cantiello, et al., \emph{A\&A} \textbf{499} 279--290 (2009).

\bibitem{muijres11}
L.~E.~Muijres, \emph{A\&A} \textbf{526}  A32$+$ (2011).

\bibitem{vink07}
J.~S.~Vink, \emph{A\&A} \textbf{469} 707--711 (2007).

\end{thebibliography}

\newpage

{\bf M. Tanaka}: Can you study the relation between PAs derived from polarization and those from the image for WRs?

{\bf JSV}: If the WR nebula has a preferred axis, the answer would be yes. However, many of the observed WR nebula actually 
appear to be spherical. 
Furthermore, the total number of line-effect WR stars is very small, and 
I fear that one would end up in the low-number statistics regime. 
For pre-main sequence stars, the correlation between our polarization PAs and those derived from 
images has been very encouraging.\\ 

{\bf J. Brown}: Regarding clump formation sites, remember (Brown's Theorem! Isle aux Coudres conference) that creating (smallish) clumps by
redistribution (e.g. condensation) of wind gas above the photosphere does NOT introduce any polarization. This is because the clump polarization
is canceled by the cavity polarization.

{\bf JSV}: I agree, and this suggests that the clumps cannot form in the wind, but they are probably already present in the photosphere.
This is consistent with recent clump-formation scenarios that invoke the Fe opacity bump (Cantiello et al. 2009).\\ 

{\bf J. Cassinelli}: Again in regards to the polarization-Z connection for Wolf-Rayet stars, the higher the Z the larger the M-dot. If a high-Z WR
star has a disk it would also have a strong polar wind. As shown by Taylor \& Cassinelli (1992), there can be strong depolarization caused by the
polar component of the wind. E.g., if the polar ejecta has an M-dot of only 1/10 that in the equator, it will cancel out the polarization
completely. So physical geometry cancellation is also an important aspect of the total polarization.

{\bf JSV}: Yes geometric cancellation can certainly occur, and the absence of polarization should not be taken as evidence for absence of 
asymmetry, but only as absence of evidence.\\ 

{\bf K. Nordsieck}: On Wolf-Rayets, it has always struck me that the ones that are polarized are very polarized, and there is no in between. I
suspect this is because the WR class is so heterogeneous. Maybe all WR stars of a particular sub-class are polarized.

{\bf JSV}: It is true that the WR class is very heterogeneous, but it does not seem the be the case that objects from 
one particular sub-class are  predominantly polarized, whilst members from other sub-classes are not. 
Still, until now we have only been able to study 
small samples, and it would thus make little sense to divide them up any further into sub-groups (such as WN/WC, early/late, single/binary, etc.). 
Hopefully, it will become possible in the near future to study significantly large sub-samples! \\

{\bf J. Hoffman}: Have you seen any of the work on AGN/Seyfert galaxy line polarization? Some of these objects show similar line effects (e.g. Smith
et al. 2005). Q-U loops also show up in many supernovae.

{\bf JSV}: Yes, Smith et al. (2005) found $QU$ loops in Seyfert galaxies, which may be related to rotating accretion disks. 
For the $QU$ loops in SNe one probably needs to consider alternative explanations.\\

{\bf D. Leonard}: So, WR stars with Z > 0.5 Zsun are found to have spherical outflows. What's the latest on the metallicity of host galaxies of GRBs?

{\bf JSV}: Whilst a low metallicity host is not an {\it absolute} requirement, there appears to be a strong 
preference for long GRBs to occur in low metallicity host galaxies, at least as far as I am aware (but my 
extragalactic knowledge is limited!). \\

{\bf R. Ignace}: In principle, the strong forbidden lines of WR stars can help constrain geometry. Spherical winds produce flat-top lines from the
large scale wind, but deviations from flat-top suggest non-spherical flow. WR 134 is an example that is polarized and also has somewhat
double-horned emission lines.

{\bf JSV}: I agree, and I think it is extremely exciting and worthwhile to look for correlations 
between Stokes $QU$ polarization versus Stokes $I$ intensity indicators!

\end{document}